\documentclass[conference]{IEEEtran}
\IEEEoverridecommandlockouts
\usepackage{amsmath,amssymb,amsfonts}
\usepackage{algorithm}
\usepackage{graphicx}
\usepackage{textcomp}
\usepackage{xcolor}
\usepackage{algpseudocode}
\usepackage[hidelinks]{hyperref}
\usepackage{url}
\usepackage[sort&compress, numbers]{natbib}
\usepackage{color}
\usepackage{bm}
\usepackage{makecell}
\usepackage{tabu, booktabs}
\usepackage{diagbox}

\def\BibTeX{{\rm B\kern-.05em{\sc i\kern-.025em b}\kern-.08em
    T\kern-.1667em\lower.7ex\hbox{E}\kern-.125emX}}
\begin{document}

\title{Community Battery Energy Storage Systems for Enhancing Distribution System Operation: \\A Multi-objective Optimization Approach
\thanks{This work was supported in part by the Australian Research Council (ARC) Discovery Early Career Researcher Award (DECRA) under Grant DE230100046.}
}

\author{\IEEEauthorblockN{Yunqi Wang\textsuperscript{1,2}, Hao Wang\textsuperscript{1,2*}, Markus Wagner\textsuperscript{1,2}, Ariel Liebman\textsuperscript{1,2}}\\
\IEEEauthorblockA{\textsuperscript{1}Department of Data Science and AI, Faculty of IT, Monash University, Australia \\
\textsuperscript{2}Monash Energy Institute, Monash University, Australia\\
Email: \texttt{\{yunqi.wang,hao.wang2,markus.wagner,ariel.liebman\}@monash.edu}}
\thanks{*Corresponding author: Hao Wang.}
}

\maketitle

\begin{abstract}
The growing penetration of distributed energy resources (DERs) in distribution networks (DNs) raises new operational challenges, particularly in terms of reliability and voltage regulation. In response to these challenges, we introduce an innovative DN operation framework with multi-objective optimization, leveraging community battery energy storage systems (C-BESS). The proposed framework targets two key operational objectives: first, to minimize voltage deviation, which is a concern for a distribution network service provider (DNSP), and second, to maximize the utilization of DERs on the demand side. Recognizing the conflicting nature of these objectives, we utilize C-BESS to enhance the system's adaptability to dynamically adjust DN operations. The multi-objective optimization problem is solved using the non-dominated sorting genetic algorithm-II (NSGA-II). Case studies using real-world data are conducted to validate the effectiveness of the proposed framework. The results show significant improvements in voltage regulation and DER utilization, demonstrating the potential of C-BESS in enabling more reliable DN operation. Our findings contribute to the ongoing discourse on the role of C-BESS in DN operation enhancement and DER integration.
\end{abstract}


\section{Introduction}

The increasing penetration of distributed energy resources (DERs) in power systems has provided unique opportunities and challenges for distribution network service providers (DNSPs)~\cite{hu2017transactive}. On the one hand, DERs, especially renewable energy sources (RESes), play a significant role in reducing greenhouse gas emissions and enhancing energy sustainability. On the other hand, the intermittent and uncertain nature of renewable DERs poses significant challenges to the stability, reliability, and operation of power systems~\cite{quint2019transformation}.

The management of a distribution system with a high penetration of DERs invariably entails a complex interplay of conflicting objectives, thus requiring sophisticated optimization methods such as multi-objective optimization. Many recent studies have delved into this subject. For instance, the study in \cite{ren2010multi} proposed a multi-objective optimization for the operation of distributed energy storage systems, with a focus on economic and environmental aspects. The study in~\cite{sousa2015multi} focused on multi-objective optimization aiming to minimize the cost of all DERs and minimize the voltage magnitude difference, while coordinating DERs considering both active power and reactive power. Lu \textit{et al.}~\cite{lu2022multi} developed a multi-objective optimization model for a microgrid system, considering the trade-off among the profit of the power company, peak-valley load difference, and the user's satisfaction after participating in the demand response. Ahmadi \textit{et al.}~\cite{ahmadi2021multi} optimized the trade-off between maximizing renewable energy utilization and minimizing the cost of grid support services.

However, most of the above studies primarily focused on the DERs, such as renewable units. There has been little attention given to the role of battery energy storage systems (BESS) to harness their flexibility in DN operations. In recent years, BESS, in particular on a community scale, has started to gain traction as a promising response to challenges faced by DN operations. Compared to distributed behind-the-meter batteries, community BESS (C-BESS), often functioning as a centralized energy storage system, can offer the advantage of orchestrating the diverse energy demands of consumers and local supplies within the community~\cite{kalkbrenner2019residential,zhao2020virtual}. C-BESS can facilitate better energy management by allowing a larger pool of energy to be stored and distributed as needed, thereby reducing instances of energy wastage. Rahbar \textit{et al.}~\cite{rahbar2014real} presented a real-time community energy storage management strategy for integrating renewable energy into a microgrid. Tomin \textit{et al.}~\cite{tomin2022design} discussed community-level energy management in microgrids using demand response and distributed storage for wind power generation. Liu \textit{et al.}~\cite{10003636} investigated the benefits of C-BESS with shared energy storage. Tian \textit{et al.}~\cite{tian2015hierarchical} proposed a two-level hierarchical optimization method for the microgrid community's energy management system in a smart grid. Although the aforementioned studies have advanced the understanding of optimizing C-BESS operations in power systems with high DER penetration, they mainly emphasized economic performance. The technical dimension concerning reliable operation and voltage regulation has been understudied, and the potential conflicts between economic and technical metrics were often overlooked when using single-objective optimization approaches. In terms of technical objectives, voltage deviation minimization and DER utilization maximization are of prime concern for the reliable operation of DNs with DERs~\cite{ghadi2019review}. These two objectives are often in conflict and thus needed to be comprehensively investigated. It is also crucial to understand the value of C-BESS in mediating such conflicting objectives, leading to a win-win situation.

The above research gaps motivate our study. We present a novel multi-objective optimization framework for the reliable operation of DNs using C-BESS as a flexible resource. The proposed framework considers the conflicting objectives of voltage deviation minimization and DER utilization maximization and demonstrates the value of C-BESS in enhancing DN operations.
The contributions of this paper are as follows.
\begin{enumerate}
\item We propose a multi-objective optimization framework for DN operations, which captures two stakeholders' interests, including DNSP's voltage variance minimization and energy users' DER utilization maximization. This leads to a clear understanding of the trade-off between both stakeholders' interests in the DN operation.

\item We utilize C-BESS taking on a pivotal role, not just as an auxiliary energy source but as a flexible participant in the DN operation, demonstrating C-BESS's value in improving the trade-off between two objectives and thus enhancing the DN operation.   

\item Through case studies using real-world data, we demonstrate that our proposed framework using C-BESS not only improves voltage regulation, but also increases DER utilization.  
\end{enumerate}

\section{Voltage Variation in DN with DER} \label{Section2}

\begin{figure}[!tp]
\centering
\includegraphics[width=6.4cm]{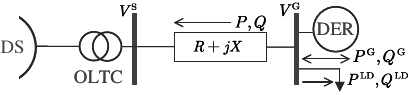}
\caption{Sample of a two-bus system with user side installed with a DER}
\label{Fig.1}
\end{figure}

When a distributed generator is connected to a DN, it needs to be operated at a higher voltage to export power, compared to the buses (e.g., PQ load) which it supplies power. This impacts the power flow and relevant voltage profiles in the system through feeders. Taking a two-bus DN as an example, assuming the buses are connected via an overhead line with the impedance of $R + jX$, the relationship between the sending-end voltage $V^{\rm S}$ and receiving-end voltage $V^{\rm R}$ can be approximated as 
 \begin{align}
    V^{\rm R} \approx V^{\rm S}+R P+X Q, \label{1}
\end{align}
where $P$ and $Q$ denote active power and reactive power flow. In addition, given a consumer with a DER, as shown in Fig.~\ref{Fig.1}, the direction of the power flow from a consumer with DER is reversed compared to the flow to the pure load, and the voltage at the point of connection of the DER is higher than the sending-end voltage. If the DER is connected where the voltage is $V^{\rm G}$, it can generate $P^{\rm G}$ and $Q^{\rm G}$ active and reactive power, and the net active and reactive load are denoted by $P^{\rm {LD}}$ and $Q^{\rm {LD}}$. Based on \eqref{1}, the voltage variation along the DN can be written as 
\begin{align}
    \!\!\!\!\Delta V=V^{\rm{G}}-V^{\rm{S}} \approx \frac{R\left(P^{\rm G}-P^{\rm {LD}}\right)+X\left(\pm Q^{\rm G} - Q^{\rm {LD}}\right)}{V^{\rm{G}}}. \!\!\label{2}
\end{align}

DERs consistently export active power, represented as $+P^{\rm G}$. Moreover, they may either export or import reactive power, denoted as $\pm Q^{\rm G}$. This variability of reactive power flow depends heavily on the inherent characteristics of the particular DERs connected to the DN. In contrast, load profiles in the network consume both active power and reactive power, represented by $-P^{\rm {LD}}$ and $-Q^{\rm {LD}}$, respectively. In these expressions, negative and positive signs precisely denote the direction of power flow at a bus, either injection or ejection. For simplicity in the following sections, the sign distinctions for power flow will be omitted, with the understanding that the symbol embraces both negative and positive values.  


\section{Proposed Operation Framework}

\subsection{Synopsis Description}
\begin{figure}[!bp]
\centering
\includegraphics[width=8.4cm]{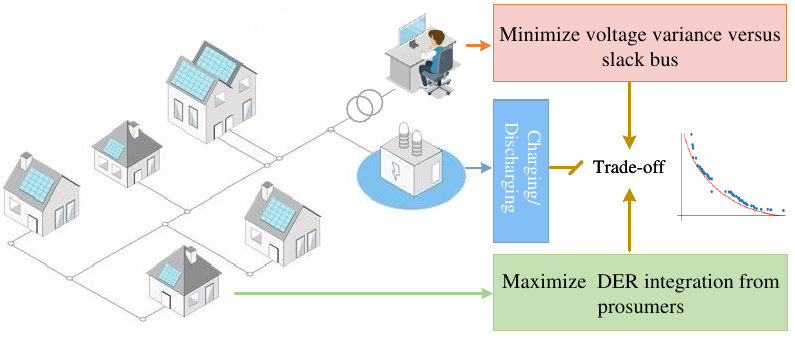}
\caption{Schematic of system structure and the proposed operation framework. We assume that the capacity of the C-BESS is capable to support the local community's energy consumption.}
\label{Fig.2}
\end{figure}
The depicted local distribution network, as shown on the left-hand side of Fig.~\ref{Fig.2}, comprises users and a C-BESS. For clarity, the users, as discussed in Section~\ref{Section2}, are divided into two categories: prosumers, who have installed DERs, and consumers, who are purely energy users. From the perspective of prosumers, they aim to maximize the consumption and export of their DER output. However, there exists a clear contrast between the interests of prosumer-desired behind-the-meter devices and the DNSP to minimize voltage variance. To handle this conflict of interests, we apply multi-objective optimization to explore the trade-off between two objectives of prosumers and the DNSP. The outcome of this non-dominated trade-off offers valuable insights, guiding decision-making concerning the preference balance between voltage variance and DER utilization. In particular, we propose to use a C-BESS, tasked with regulating energy flow through strategic charging and discharging, to mitigate this conflict. In detail, C-BESS charges during periods of excess DER production, thereby supporting the prosumers’ objective of maximizing DER utilization. Concurrently, by storing this excess energy, it prevents potential voltage spikes, thus addressing the DNSP’s aim of voltage variance minimization. The operational framework for this concept is given on the right-hand side of Fig.~\ref{Fig.2}. 
We primarily focus on the centralized dispatch of the C-BESS's active power to aid in voltage deviation control. The support from the C-BESS can effectively harmonize the disparate goals of the DNSP and prosumers.

\subsection{The BESS Model}
The BESS can store and release energy through charging and discharging behaviors. The simplified mathematical models of BESS can be expressed below: 
\begin{subequations}
\begin{gather}
\begin{array}{l}
\Phi _{b,t+1}^{\mathrm{}}=\Phi _{b,t}^{\mathrm{}}+P_{b,t}^{\mathrm{Chg}}\cdot\eta _b\cdot \Delta t-\dfrac{P_{b,t}^{\mathrm{Dis}}\cdot \Delta t}{\eta _b}, \\
-\Phi _{b,t}^{\mathrm{}}\cdot \Delta t\cdot l^{\mathrm{}},\quad\forall b,t, 
\end{array} \label{b1}\\
S O C_{i, t}=\frac{\Phi_{b, t}^{\rm{}}}{\Phi_b^{\rm{Rate}}}, \quad\forall b, t,\label{b2}\\
\underline{S O C}_{b} \leqslant S O C_{b, t} \leqslant \overline{S O C}_b, \quad \forall b,t, \label{b3}\\
\left\{ \begin{array}{l}
	0\leqslant P_{b,t}^{\rm{Chg}}\leqslant   \overline P_b^{\rm{BESS}}\mu _{b,t}\\
	0\leqslant P_{b,t}^{\rm{Dis}}\leqslant   \overline P_b^{\rm{BESS}}(1-\mu _{b,t})\\
\end{array} \right., \forall b,t \label{b4}
\\
\Phi _{b,0}= \Phi _b^{\rm Init}{\rm{; }}\quad\Phi _{b,T}^{\rm } \geqslant \Phi _b^{\rm End}, \label{b5}
\end{gather}
\end{subequations}
where $\eta_b$ denotes the charging efficiency of battery $b$. Normally, there is a reciprocal relationship between the charging and discharging efficiency; $\Phi _{b,t + 1}^{\rm }$ and $\Phi _{b,t}^{\rm }$  denote the battery energy at time $t$ and $t+1$, respectively; \(\Phi_b^{\rm Rate}\) refers to the rated capacity of the battery; \(P_{b,t}^{\rm Chg}\) and \(P_{b,t}^{\rm Dis}\) are charging power and discharging power of the C-BESS; $SOC_{b,t}$ denotes the state of charge of BESS at time $t$; \(l^{\rm }\) is leakage loss coefficient, representing the energy-based degradation cost; $\mu _{b,t}$ is binary variable identifying the battery states; $\Phi_b^{\rm Init}$ and $\Phi_b^{\rm End}$ impose the initial and end energy for battery $b$; \(\underline {\left(  \bullet  \right)} and \overline {\left(  \bullet \right)} \) denote the lower- and upper-bounds.  

Equation \eqref{b1} imposes the energy balance of the battery including energy difference, energy losses during charging or discharging, and leakage loss; equations \eqref{b2}-\eqref{b3} describe the \(SOC\) constraints and $SOC$ limitation; \eqref{b4} prevents simultaneous charging and discharging; \eqref{b5}
defines the initial and final energy stored state.

\subsection{Modeling C-BESS Operation in a DN}

The C-BESS is considered as a centralized and controllable component within the DN. As the size of the C-BESS is considerable compared to the DN, it can affect the power flow and the operation of the DN, such that it can help manage the power flow and ensure system reliability. The charging and discharging characteristics of the C-BESS need to be modeled and incorporated effectively. Mathematically, the C-BESS in a DN can be modeled as \eqref{s1}-\eqref{s7}. Note that a linearized Dist-flow in~\cite{zhang2016robust} is utilized to conduct power flow analysis considering the common radial topology of DNs, to reduce the complexity of formulation and computation. 
\begin{subequations}
\begin{gather}
\left\{\begin{array}{l}
P_{i+1, t}=P_{i, t}+P_{i+1, t}^{\rm G}-P_{i+1, t}^{\rm {LD}}-P_{i+1, t}^{\rm {Lat}}-P_{i, t}^{\rm{BESS}} \\	
Q_{i+1, t}=Q_{i, t}-Q_{i+1, t}^{\rm {LD}}-Q_{i+1, t}^{\rm {Lat}}, \quad\forall i, t,  
\end{array} \right. \label{s1}
\\
P_{i, t}^{\rm{BESS}} = {P}_{i,t}^{\rm{Chg}} - {P}_{i,t}^{\rm{Dis}}, \quad \forall i, t, \label{s2}\\
P_{i,t}^{\mathrm G} =\left\{ \begin{array}{l}
	P_t^{\mathrm {Grid}} + P_{i,t}^{\rm {DER}},  \quad \forall t, i=1\\
	P_{i,t}^{\mathrm {DER}}, \quad \forall t, i\neq 1,
\end{array} \right. \label{s3}\\
\underline P^{\mathrm{Grid}} \leqslant P^{\mathrm{Grid}}_t \leqslant \overline P^{\mathrm{Grid}}, \quad \forall t, \label{s4}
\\
\underline P^{\rm{DER}}_i\leqslant P^{\rm{DER}}_{i, t} \leqslant \overline P^{\rm{DER}}_i, \quad \forall i,t, \label{s5} \\
V_{i+1, t} = V_{i,t} - \frac{\left(R_iP_{i,t} + X_iQ_{i,t}\right)}{V_0}, \quad \forall i,t, \label{s6}\\
\underline V_i\leqslant V_{i, t} \leqslant \overline V_i, \quad \forall i,t. \label{s7}
\end{gather}
\end{subequations}

Herein, $P_{i,t}$ and $Q_{i,t}$ denote active and reactive power flow passing the main branch through bus $i$ at time $t$, respectively; $P_{i,t}^{\rm Lat}$ and $Q_{i,t}^{\rm Lat}$ represent the active and reactive power flow passing the lateral branch through bus $i$ at time $t$, respectively; $P_t^{\rm {Grid}}$ denotes the power exported from the bulk grid; $P_{i,t}^{\rm {DER}}$ denotes the DER power available to be utilized from prosumer at bus $i$. We mainly focus on the solar power from PV panels as DERs, and the upper bound of DER's output corresponds to the maximum power generation of a PV panel. 

Equation \eqref{s1} is the power balance modified upon linearized Dist-flow; Equation \eqref{s2} gives the intermediate variable of describing the difference between charging and discharging power of battery; Equation \eqref{s3} describes the nodal power injection from different types of energy sources (e.g., bulk grid); Equations \eqref{s4}-\eqref{s5} impose the power limitation of different types of energy sources; Equation \eqref{s6} denotes the voltage relationship in the large-scale network extended from Equation \eqref{2}. It is noted that our current approximation does not account for power loss during delivery. Besides, there are some shunt compensators installed in DNs, which may contribute to the supply or absorption of reactive power, these parts are outside the scope of this paper and thus have not been incorporated into our current model. Moreover, the capacity for reactive power generation or absorption from DERs hinges on the specific DERs deployed within the DN. While our study presents a generalized approach, we acknowledge the potential for more detailed modeling considering the variety of DERs and their respective capabilities. Equation \eqref{s7} keeps the voltage within the safety range. 

\subsection{Multi-objective Optimization Formulation}

The explicit formulations of the proposed framework of the multi-objective optimization problem can be modeled as \eqref{3} and \eqref{4} for a DN with a set of buses $\mathcal{N}$ in a temporal horizon set $\mathcal{T}$. In detail, equation \eqref{3} minimizes the square value of normalized voltage variance with slack bus voltage $V_0$ being reference, and \eqref{4}  is directed towards maximizing the total energy consumption served by DER, an objective that aligns with the interests of prosumers who seek to maximize the usage of their DER installations. 
\begin{align}
& \min f_1= \displaystyle\sum_{t\in \mathcal{T}} \displaystyle\sum_{i\in \mathcal{N}} {\left( \frac{V_{i,t}-V_0}{\overline{V}_i-\underline{V}_i} \right)}^2, \label{3}\\
&\max f_2 = \displaystyle\sum_{t\in \mathcal{T}} \displaystyle\sum_{i\in \mathcal{N}}P_{i,t}^{\rm {DER}}. \label{4}
\end{align}

The objectives are subject to constraints \eqref{b1}-\eqref{b5} and \eqref{s1}-\eqref{s7}. Considering a C-BESS which is installed on a bus individually, the battery index $b$ and bus index $i$ can be consolidated. All the constraints construct a solution space, such that every time the solver produces a solution, it is instantly classified as feasible or infeasible, enabling the solver to converge on a feasible solution space.

Within this optimization problem, the decision variables constitute an array of elements that play a crucial role in determining the most optimal solution. These decision variables include the vector of voltage levels at different buses, $\bm V$, the output of DERs, $\bm{P}^{\mathrm{DER}}$, and the charging and discharging power of the community batteries, represented by $\bm{P}^{\mathrm{Chg}}$ and $\bm{P}^{\mathrm{Dis}}$ respectively. Additionally, variables relevant to power flow are also incorporated into this optimization framework. Specifically, by carefully managing $\bm{P}^{\mathrm{Chg}}$ and $\bm{P}^{\mathrm{Dis}}$, it is possible to further minimize voltage deviations and enhance the utilization of distributed renewable energy, thus fulfilling the two objectives of the proposed optimization framework.

In solving the complexities of the problem presented in this study, we turn to the field of evolutionary multi-objective optimization. As a subfield of evolutionary computation, evolutionary multi-objective optimization is notable for its utilization of population-based, meta-heuristic approaches. These approaches are designed to find multiple optimal solutions in a single run, which is especially beneficial when dealing with complex, multi-objective problems where numerous conflicting objectives must be considered simultaneously. The field of evolutionary multi-objective optimization includes a variety of algorithms, such as the Non-dominated Sorting Genetic Algorithm II (NSGA-II)~\cite{deb2002fast}, Strength Pareto Evolutionary Algorithm 2 (SPEA-2)~\cite{zitzler2001spea2}, and Multi-objective Particle Swarm Optimization (MOPSO)~\cite{coello2007evolutionary}. Each of these algorithms offers unique capabilities to manage complex optimization challenges, with specific strengths and trade-offs that make them more or less suitable for different problem domains.

Our chosen algorithms for this research are the popular NSGA-II and SPEA-2. Both fall into the category of population-based optimization heuristics: both use non-dominated sorting processes and other techniques to navigate the trade-off space between multiple conflicting objectives. Moreover, both typically employ unary and binary variation operators to create new solutions based on those observed so far. NSGA-II operates by using the Pareto dominance principle to evaluate and rank solutions in the objective space, successively forming and removing fronts from the initial population until the entire population is distributed into multiple rank-based fronts. This method ensures the preservation of elite solutions throughout the algorithm's execution, thereby aiming to deliver a balanced and comprehensive set of trade-off solutions. SPEA-2 operates comparably in the sense that it also uses crowding distance, not just for selection but mainly for calculating the fitness of solutions in the archive, which influences their chances of survival and inclusion in the next generation. We employ both to reduce the potential bias when interpreting solutions.

\section{Case Study}

\subsection{Experimental Setups}
\begin{figure}[!bp]
\centering
\includegraphics[width=8.4cm]{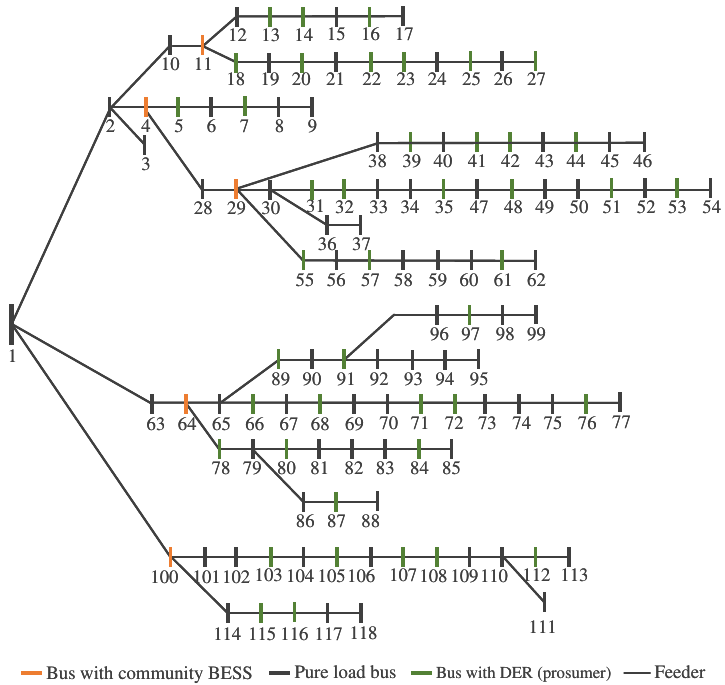}
\caption{Single line diagram of modified IEEE 118-bus DN}
\label{Fig.3}
\end{figure}

The performance of the proposed operation framework is evaluated on an IEEE standard 118-bus DN~\cite{zhang2007improved}. Five C-BESS are installed in the buses connected with the main branches of the DN. In detail, Bus 4, 11, 29, 64, and 100 correspond to C-BESS 1 to 5. The system has a peak load of 22,709.7kW and 17,041.1kvar, and the ratio between prosumer and consumer is 40\%. We set the temporal horizon in one day with a one-hour step, which is a common power system operation problem, as it also matches the day-ahead market. Moreover, the prosumer PV data is obtained from Pecanstreet~\cite{Precanstreet}, and load profiles are obtained from~\cite{zhang2007improved}, then forecasted based on real-world consumption patterns which are customized upon original data with the data acquired from AEMO \cite{AEMO}. Two cases are developed to test the effectiveness of the utilization of C-BESS. The experiment is conducted on Python platform with the help of Pymoo~\cite{blank2020pymoo} on a PC with Intel(R) Core(TM) i7-8700 CPU @ 3.20GHz with 16GB RAM. 

\textit{Case 1}: Multi-objective operation framework \textit{without} C-BESS participation.

\textit{Case 2}: Multi-objective operation framework \textit{with} C-BESS participation.

\subsection{Experimental Results}
\begin{figure}[!tp]
\centering
\includegraphics[width=7.4cm]{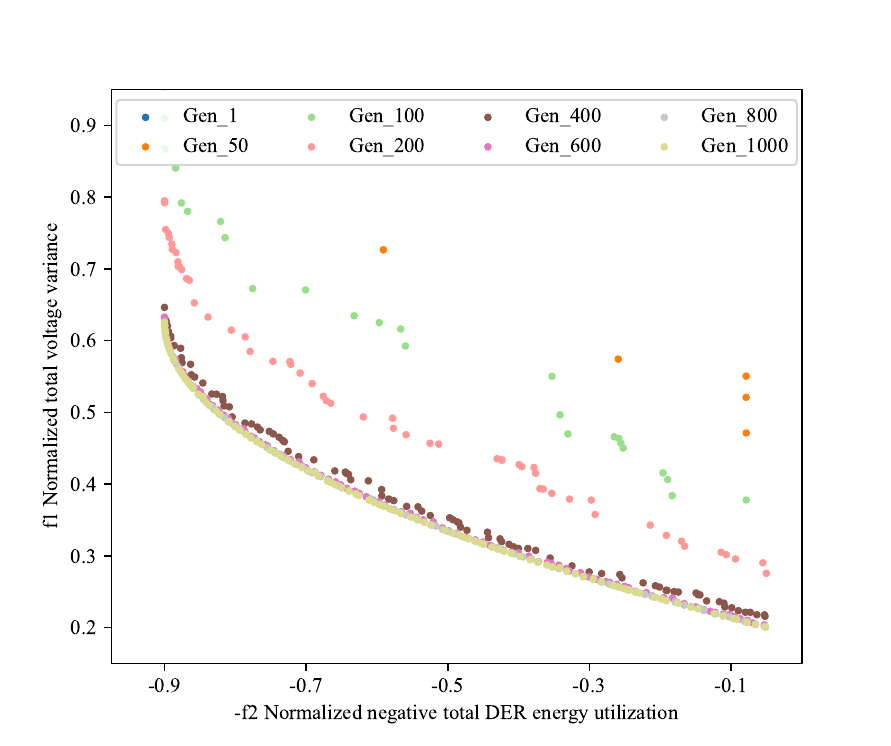}
\caption{Non-dominated solutions obtained in different generations (one run of NSGA-II). The first generation (Gen${_1}$) is not visible as it is outside of the shown ranges. In this run, 800 to 1000 generations were needed for the algorithm to converge.}
\label{Fig.4}
\end{figure}


The comparative analysis of non-dominated solutions achieved through the NSGA-II algorithm across diverse generations for Case 2 is shown in Fig.~\ref{Fig.4}, in which the objective values are normalized to facilitate an unbiased comparison. A trend of progressive improvement is distinctly noticeable as we traverse the generations. With each iteration, not only are the values of non-dominated solutions decreasing, pointing to an effective refining and honing process toward optimal solutions, but we also observe growing stability in the solutions obtained. As the number of generations increases, the solutions appear to converge, indicating that the algorithm is approaching a region of optimal solutions. This progressive stabilization also implies that several parameters integral to the problem at hand are being optimized and can subsequently be fixed. Besides, a single run with 1000 generations takes about 4 minutes.  

\begin{figure}[!bp]
\flushleft
\includegraphics[width=8.6cm]{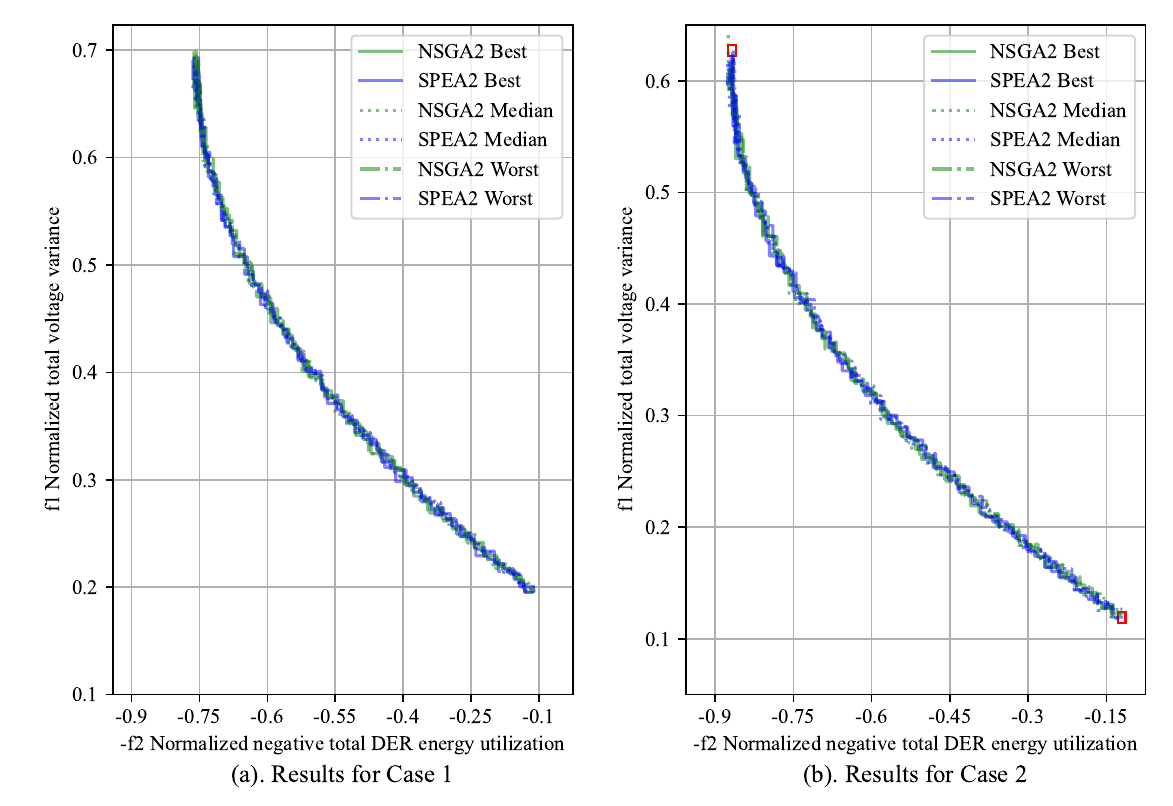}\vspace{-2mm}
\caption{Attainment surfaces for different cases using different algorithms (Conducted over 30 runs each).}
\label{Fig.5}
\end{figure}

To achieve a clearer perspective of the optimization landscape, attainment surfaces are employed. It can also provide valuable insights into the trade-offs and compromises between conflicting objectives. Moreover, in pursuit of a more holistic understanding of our problem domain, we incorporate the SPEA-2 algorithm alongside NSGA-II. Both algorithms are subjected to 30 independent runs with the same parameters set, ensuring a robust and consistent comparison between their performance metrics. The result of the attainment surfaces is given in Fig.\ref{Fig.5}. A salient observation from this figure is the consistency in the surfaces across 30 runs. The ``best" and ``worst" attainment surfaces, as determined by the hypervolume values, serve as the extremities, while the median surface, derived from the central tendency of the 30 runs, reveals a close alignment between the two algorithms. Notably, NSGA-II appears to be slightly better than SPEA-2 for our problem.

\begin{table}[!tp]
\caption{Voltage profile analysis for different cases}
\centering
\setlength{\tabcolsep}{3mm}
\begin{tabular}{|c|c|c|c|}
\hline
\diagbox{Case}{Item}& Mean & STD & Median \\
\hline
\makecell*[c]{Case 1} &1.058 & 0.009 & 1.057\\
\hline
\makecell*[c]{Case 2} &1.021 & 0.004 & 1.017\\
\hline
\end{tabular}
\label{tab1}
\end{table}

When we pivot our focus to compare Case 1 and Case 2, a significant difference in the attainment surface is observed. In Case 2, objective values are distinctly lower than those in Case 1. Considering that our problem aims to minimize the objectives, lower objective function values in Case 2 indicate an improved approximation of the Pareto frontier. It demonstrates that the introduction of the C-BESS has a positive impact on both objectives. It contributes to minimizing voltage changes and simultaneously increasing the utilization rate of DERs. Table~\ref{tab1} presents a comparative analysis of voltage profiles for different cases, key statistical measures such as the mean, standard deviation (STD), and median are compared, and values are in per-unit (p.u.). It can be seen that despite the analogical voltage profiles in Case 1 showing there is no large voltage fluctuation, Case 2 appears to have a slight edge in maintaining consistent voltage levels. Compared to Case 1, Case 2 has a slightly lower mean voltage value of 1.021. Moreover, the STD in Case 2, valued at 0.004, is considerably smaller, suggesting a more compact distribution of voltage values around the mean. This points to extremely stable voltage levels in Case 2, with minimal deviations from the average voltage level. Further, the closeness of the median value of 1.017 to the mean, affirms the high level of voltage stability in Case 2. The small difference between the mean and the median underlines a symmetric distribution of voltage values around the mean and further reinforces the notion of consistent voltage levels. 

\begin{figure}[!tp]
\flushleft 
\includegraphics[width=8.4cm,trim={0 211 0 0},clip]{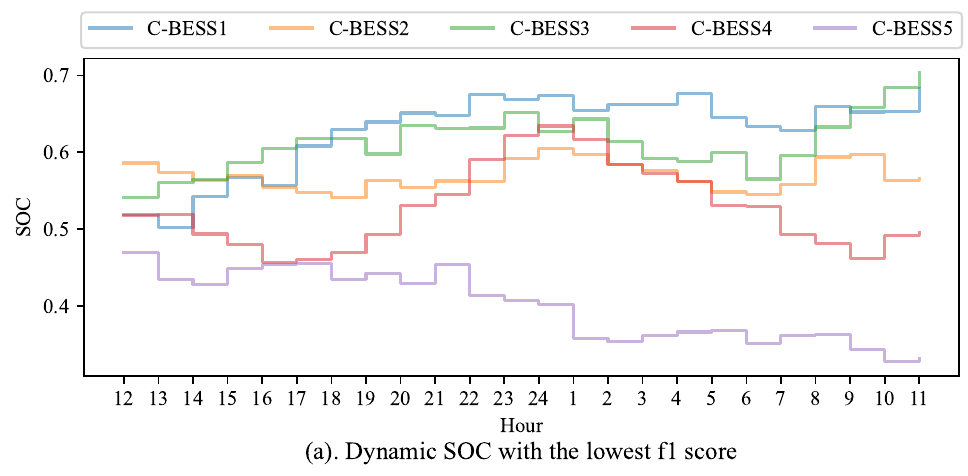}\\
\includegraphics[width=8.4cm,trim={0 0 0 224},clip]{SOC.pdf}\vspace{-2mm}
\caption{SOC for batteries of different Non-dominated solutions in Case 2.}
\label{Fig.6}
\end{figure}

Within both scenarios, the SOC dynamics of each C-BESS delineate distinct behavioral trajectories, emphasizing their discrete contributions towards the attainment of the overarching objective. Such dynamics diverge significantly from conventional electricity arbitrage behaviors, shedding light on the C-BESS's paramount role within the DN's intricate multi-objective optimization framework.

In the context depicted by subfigure (a), where the minimization of voltage deviation ($f_1$) stands paramount, there is an observable controlled oscillation in battery charging and discharging behaviors. This is demonstrative of an equilibrium in the grid's power flow and is conducive to preserving grid voltage stability. To illustrate, the SOC of C-BESS 1 initiates at 0.52 and, through a discernible trajectory, culminates at 0.68, an outcome possibly resultant from either a strategic charging approach or a decline in communal energy demand. Conversely, another battery exhibits an SOC commencement at 0.59 with a modest descent to 0.57, which could be emblematic of a day marked by normative energy consumption devoid of aggressive charging interventions.

Subfigure (b), conversely, accentuates the maximization of DER utilization ($f_2$). Herein, there is a more expansive fluctuation in the SOC, indicative of a more vigorous battery utilization strategy. This is exemplified by C-BESS 1, which, while commencing with a similar SOC as in the former scenario, diminishes to 0.47, potentially signaling an augmented community energy consumption or strategic intermittent charging. Such variances suggest these batteries' instrumental roles in bolstering an elevated DER utilization, which may be integral to assimilating renewable energy resources within the grid's infrastructure.

\section{Conclusion}

This study introduced a unique perspective on the operation of DNs amidst the growing integration of DERs. We proposed an innovative multi-objective operational framework incorporating the C-BESS, providing a viable approach to navigating the operational complexities introduced by the influx of DERs. One of the pivotal findings from this study highlights the role of C-BESS as effective resource in mitigating the conflicting nature of operational objectives. Through case studies using real-world data, we demonstrated how the C-BESS can significantly enhance voltage regulation while simultaneously boosting DER utilization. This contributes to a more adaptable and reliable operation of DNs, enabling a critical shift in the discourse surrounding DN operations, particularly in light of the challenges and opportunities posed by the integration of DERs. As part of future work, we will consider more accurate power flow analysis, moving beyond linearized Dist-flow and taking into account the power loss.

\renewcommand{\bibfont}{\footnotesize}
\bibliographystyle{IEEEtran}
\bibliography{reference.bib}

\begin{thebibliography}{10}
\providecommand{\url}[1]{#1}
\csname url@samestyle\endcsname
\providecommand{\newblock}{\relax}
\providecommand{\bibinfo}[2]{#2}
\providecommand{\BIBentrySTDinterwordspacing}{\spaceskip=0pt\relax}
\providecommand{\BIBentryALTinterwordstretchfactor}{4}
\providecommand{\BIBentryALTinterwordspacing}{\spaceskip=\fontdimen2\font plus
\BIBentryALTinterwordstretchfactor\fontdimen3\font minus
  \fontdimen4\font\relax}
\providecommand{\BIBforeignlanguage}[2]{{%
\expandafter\ifx\csname l@#1\endcsname\relax
\typeout{** WARNING: IEEEtran.bst: No hyphenation pattern has been}%
\typeout{** loaded for the language `#1'. Using the pattern for}%
\typeout{** the default language instead.}%
\else
\language=\csname l@#1\endcsname
\fi
#2}}
\providecommand{\BIBdecl}{\relax}
\BIBdecl

\bibitem{hu2017transactive}
J.~Hu, G.~Yang, K.~Kok, Y.~Xue, and H.~W. Bindner, ``Transactive control: a
  framework for operating power systems characterized by high penetration of
  distributed energy resources,'' \emph{Journal of Modern Power Systems and
  Clean Energy}, vol.~5, no.~3, pp. 451--464, 2017.

\bibitem{quint2019transformation}
R.~Quint, L.~Dangelmaier, I.~Green, D.~Edelson, V.~Ganugula, R.~Kaneshiro,
  J.~Pigeon, B.~Quaintance, J.~Riesz, and N.~Stringer, ``Transformation of the
  grid: The impact of distributed energy resources on bulk power systems,''
  \emph{IEEE Power and Energy Magazine}, vol.~17, no.~6, pp. 35--45, 2019.

\bibitem{ren2010multi}
H.~Ren, W.~Zhou, K.~Nakagami, W.~Gao, and Q.~Wu, ``Multi-objective optimization
  for the operation of distributed energy systems considering economic and
  environmental aspects,'' \emph{Applied Energy}, vol.~87, no.~12, pp.
  3642--3651, 2010.

\bibitem{sousa2015multi}
T.~Sousa, H.~Morais, Z.~Vale, and R.~Castro, ``A multi-objective optimization
  of the active and reactive resource scheduling at a distribution level in a
  smart grid context,'' \emph{Energy}, vol.~85, pp. 236--250, 2015.

\bibitem{lu2022multi}
Q.~Lu and Y.~Zhang, ``A multi-objective optimization model considering users'
  satisfaction and multi-type demand response in dynamic electricity price,''
  \emph{Energy}, vol. 240, p. 122504, 2022.

\bibitem{ahmadi2021multi}
B.~Ahmadi, O.~Ceylan, and A.~Ozdemir, ``A multi-objective optimization
  evaluation framework for integration of distributed energy resources,''
  \emph{Journal of Energy Storage}, vol.~41, p. 103005, 2021.

\bibitem{kalkbrenner2019residential}
B.~J. Kalkbrenner, ``Residential vs. community battery storage
  systems--consumer preferences in germany,'' \emph{Energy Policy}, vol. 129,
  pp. 1355--1363, 2019.

\bibitem{zhao2020virtual}
D.~Zhao, H.~Wang, J.~Huang, and X.~Lin, ``Virtual energy storage sharing and
  capacity allocation,'' \emph{IEEE transactions on smart grid}, vol.~11,
  no.~2, pp. 1112--1123, 2020.

\bibitem{rahbar2014real}
K.~Rahbar, J.~Xu, and R.~Zhang, ``Real-time energy storage management for
  renewable integration in microgrid: An off-line optimization approach,''
  \emph{IEEE Transactions on Smart Grid}, vol.~6, no.~1, pp. 124--134, 2014.

\bibitem{tomin2022design}
N.~Tomin, V.~Shakirov, A.~Kozlov, D.~Sidorov, V.~Kurbatsky, C.~Rehtanz, and
  E.~E. Lora, ``Design and optimal energy management of community microgrids
  with flexible renewable energy sources,'' \emph{Renewable Energy}, vol. 183,
  pp. 903--921, 2022.

\bibitem{10003636}
Y.~Liu, Y.~Wang, H.~Xi, J.~Lin, and J.~Ma, ``Community energy cooperation with
  shared energy storage for economic-environment benefits,'' in \emph{2022 IEEE
  PES Innovative Smart Grid Technologies - Asia (ISGT Asia)}, 2022, pp.
  230--234.

\bibitem{tian2015hierarchical}
P.~Tian, X.~Xiao, K.~Wang, and R.~Ding, ``A hierarchical energy management
  system based on hierarchical optimization for microgrid community economic
  operation,'' \emph{IEEE Transactions on Smart Grid}, vol.~7, no.~5, pp.
  2230--2241, 2015.

\bibitem{ghadi2019review}
M.~J. Ghadi, S.~Ghavidel, A.~Rajabi, A.~Azizivahed, L.~Li, and J.~Zhang, ``A
  review on economic and technical operation of active distribution systems,''
  \emph{Renewable and Sustainable Energy Reviews}, vol. 104, pp. 38--53, 2019.

\bibitem{zhang2016robust}
C.~Zhang, Y.~Xu, Z.~Y. Dong, and J.~Ma, ``Robust operation of microgrids via
  two-stage coordinated energy storage and direct load control,'' \emph{IEEE
  Transactions on Power Systems}, vol.~32, no.~4, pp. 2858--2868, 2016.

\bibitem{deb2002fast}
K.~Deb, A.~Pratap, S.~Agarwal, and T.~Meyarivan, ``A fast and elitist
  multiobjective genetic algorithm: Nsga-ii,'' \emph{IEEE transactions on
  evolutionary computation}, vol.~6, no.~2, pp. 182--197, 2002.

\bibitem{zitzler2001spea2}
E.~Zitzler, M.~Laumanns, and L.~Thiele, ``Spea2: Improving the strength pareto
  evolutionary algorithm,'' \emph{TIK report}, vol. 103, 2001.

\bibitem{coello2007evolutionary}
C.~A.~C. Coello, \emph{Evolutionary algorithms for solving multi-objective
  problems}.\hskip 1em plus 0.5em minus 0.4em\relax Springer, 2007.

\bibitem{zhang2007improved}
D.~Zhang, Z.~Fu, and L.~Zhang, ``An improved ts algorithm for loss-minimum
  reconfiguration in large-scale distribution systems,'' \emph{Electric power
  systems research}, vol.~77, no. 5-6, pp. 685--694, 2007.

\bibitem{Precanstreet}
\BIBentryALTinterwordspacing
{Precanstreet ORG}. [Online]. Available:
  \url{https://www.pecanstreet.org/dataport/}
\BIBentrySTDinterwordspacing

\bibitem{AEMO}
\BIBentryALTinterwordspacing
{AEMO}. [Online]. Available:
  \url{https://aemo.com.au/energy-systems/electricity/national-electricity-market-nem/data-nem}
\BIBentrySTDinterwordspacing

\bibitem{blank2020pymoo}
J.~Blank and K.~Deb, ``Pymoo: Multi-objective optimization in python,''
  \emph{IEEE Access}, vol.~8, pp. 89\,497--89\,509, 2020.

\end{thebibliography}

\end{document}